# Data Mining for better material synthesis: the case of pulsed laser deposition of complex oxides

Steven R. Young[1,2], Artem Maksov[2,3,4], Maxim Ziatdinov[2,3], Ye Cao[2,3], Matthew Burch[2,3], Janakiraman Balachandran[2,3], Linglong Li[2,3,5], Suhas Somnath[2,3,6], Robert M. Patton[2,3], Sergei V. Kalinin[2,3,*] and Rama K. Vasudevan[2,3]

[1]Computational Sciences and Engineering Division, [2]Institute for Functional Imaging of Materials, and [3]Center for Nanophase Materials Sciences, Oak Ridge National Laboratory, Oak Ridge TN 37831, USA

[4]Bredesen Center for Interdisciplinary Research, University of Tennessee, Knoxville, Tennessee 37996, United States

[5]Multi-disciplinary Materials Research Center, Frontier Institute of Science and Technology, Xi'an Jiaotong University, Xi'an, Shaanxi, 710049, China

[6]Advanced Data and Workflows Group, National Center for Computational Sciences, Oak Ridge National Laboratory, Oak Ridge TN 37831, USA

---

* sergei2@ornl.gov

**Abstract**

The pursuit of more advanced electronics, and finding solutions to energy needs often hinges upon the discovery and optimization of new functional materials. However, the discovery rate of these materials is alarmingly low. Much of the information that could drive this rate higher is scattered across tens of thousands of papers in the extant literature published over several decades but is not in an indexed form, and cannot be used in entirety without substantial effort. Many of these limitations can be circumvented if the experimentalist has access to systematized collections of prior experimental procedures and results. Here, we investigate the property-processing relationship during growth of oxide films by pulsed laser deposition. To do so, we develop an enabling software tool to (1) mine the literature of relevant papers for synthesis parameters and functional properties of previously studied materials, (2) enhance the accuracy of this mining through crowd sourcing approaches, (3) create a searchable repository that will be a community-wide resource enabling material scientists to leverage this information, and (4) provide through the Jupyter notebook platform, simple machine-learning-based analysis to learn the complex interactions between growth parameters and functional properties (all data/codes available on https://github.com/ORNL-DataMatls). The results allow visualization of growth windows, trends and outliers, and which can serve as a template for analyzing the distribution of growth conditions, provide starting points for related compounds and act as feedback for first-principles calculations. Such tools will comprise an integral part of the materials design schema in the coming decade.

The major eras in human civilization are demarked by the main materials available or discovered during those time periods, being stone, bronze, or iron, with each material advance resulting in a disruptive shift in human civilization. Presently, the demand for new materials stems from a huge number of industries ranging from biomedical to information and energy. At the same time, the desire for new, better and more efficient materials is also driven by the energy requirements of a growing world, necessitating superior energy harvesting and storage solutions, along with more efficient converters and transducers to harness the stored power for applications.[1] The new materials that will lead to disruptive events in these industries will be complex, given the increasing demands made of them, and will often have to exhibit multiple coupled functionalities (e.g., a room-temperature multiferroic material[2]). The downside to this complexity is that the rate of discovery and optimization of such complex materials is slow. The discovery of most material classes is serendipitous, as illustrated by families of new high temperature superconductors[3], multiferroic materials[2,4], or spinel[5] and olivine[6] classes of battery cathode materials.

Beyond discovery, the improvement rate of such complex materials is very limited, being restricted largely to the linear optimization approach, where one compound is slowly adjusted by means of doping and characterization, which is tedious, time-consuming, and largely driven by intuition of the researcher. For example, one can compare Moore's law for Si-based devices, which predicts exponential growth, vs. the much slower growth of battery technology, which is based on significantly more complicated compounds and involves at least three distinct transport functions (as opposed to 1, for electronic circuits)[2]. Furthermore, the associated publication volume has also become too large[7], far exceeding the capability of individual domain experts to know the literature or exchange expertise with other experts. While many details and performance are published, it is impossible for people to read them, much less determine where the studies fit in the larger framework, and understand the structure-property relations (i.e., why) and the controlling factors.

A theory-based approach to address this problem is offered by the Materials Genome Initiative[8,9], which develops a computational materials science and engineering based approach towards the discovery of new functional materials. These large databases of first principles calculations can be data mined for interesting correlations among different materials[10,11], and can be used to further progress and speed up materials discovery. Yet, the issue is that even with this approach, the question remains as to which materials can actually be fabricated (i.e., thermodynamic phases calculated by density functional theory may be impossible to reach in practice), and it is usually not known under what conditions these materials can be synthesized. Very recently, databases and platforms for sharing of experimental data and methods have come into existence, including the Materials Data Facility[12,13], Citrination from Citrine informatics[14], Dark Reactions[15], Materials Innovation Network[16] and the Materials Data Curation System[17].

One approach is to mine the existing literature, to compile experimental databases of these parameters that can aid in prediction of synthesis pathways, as well as demonstrate the relationship between processing conditions and functional parameters to assist in mechanism determination or aid in materials optimization. Indeed, this approach has been trialed by Gaultois et al.[18] and Sparks et al.[19] for thermoelectrics. Similarly, de Pablo et al.[20] demonstrated a tool to determine thermodynamic parameters of a polymer, utilizing crowd-sourcing (in their case, via outreach towards the student community) to compile entries in a database for subsequent analysis.[21] In this regard, the progress by Olivetti and colleagues, in using natural language processing to mine papers for synthesis conditions for compounds, is a notable step.[22-24]

Here, we have developed a general tool that can automatically extract synthesis parameters and functional property information from a collection of papers, based on the open-source tool BRAT for text annotation, using regular expressions that can be modified for any type of synthesis method. While the work presented here focuses on pulsed laser deposition-grown oxides, the method should generalize well to many other fields. Crowd-sourcing allows distribution of the annotated papers to users, who can individually match growth conditions to functional properties, building entries in a centralized database. We show the example of growth conditions for epitaxial complex oxide thin films, given the large number of available papers on the topic spanning three decades. We further present analysis (fully available via a Jupyter notebook[25]) on the obtained database, with key highlights including the direct visualization of the narrow growth window for bismuth ferrite thin films, as well as a simple machine learning classifier to determine the $T_C$ of a manganite given some input processing parameters. The proliferation of centralized databases of experimental conditions, in combination with open source tools and analysis packages will become a major feature of the materials design schema in the coming years. [18,21,26]

**Workflow**

Our general workflow is shown in **Figure 1**. A collection of digital document identifiers (DOIs) for relevant papers are collected from a Web of Science search.[27] In practice, we may also use recommender systems[28] that compare the textual content of a set of papers of interest (manually selected), to those available in literature to semi-automate this process. We utilize a script to download the PDF version of papers, using the DOI list. Note that this requires full access to the journal and thus cannot be used if access to the journal is not available. In such instances, one may search public repositories for the paper if they are available, as an alternative. Nonetheless, the goal is to collect the papers as PDFs on the material/synthesis method of interest (in this case, pulsed laser deposition of particular oxides). The PDFs are then automatically converted to plain text format using the pdftotext tool provided with the Poppler library[29] for ingestion by the annotation tool, which then utilizes regular expressions and keywords to automatically highlight the parts of the text in each paper that correspond to processing conditions and physical characteristics or functional properties. More information on the annotation tool is provided in the next section. Effectively, the tool will automatically annotate (highlight and tag) the relevant parts of the text of each paper that describe processing conditions and functional properties, and present this annotated text form to the user. Typically, multiple growth conditions and multiple samples are noted in any individual paper, so these cannot be automatically compiled into a database without checking which sample was grown under which conditions, and which samples were measured, etc. Instead, we utilize crowd sourcing to achieve this matching. As an aside, theoretically natural language processing can be used for this effort, but one major complication is that much of the functional property information lies in figures, not text. Thus, we use crowd sourcing to check and match each grown material with its relevant properties and processing conditions, enabling the updating of a dynamic database. That is, users are shown only a small subset of the paper, which is calculated by simply displaying the lines before and after a keyword has been highlighted (if necessary, the user can also download the entire paper, or simply view all of the text with a single click). The users observe the text which was annotated, and then use this information to populate entries in the database. A script also allows 'scrubbing' of the database to clear errant entries, normalize units, and provide the database in a comma-separated values file for further analysis. To normalize the units, the script identifies the units in the database entries, and converts them to a predetermined standard unit. For example, we use mbar as the standard for growth pressures. As the last step, we provide a Jupyter notebook for basic visualization and analysis of the data in the database, with an example of creating a machine learning classifier to predict the transition temperature given a number of growth parameters of thin film $La_{1-x}(Sr,Ca)_xMnO_3$.

**Annotation tool**

After the relevant papers have been downloaded and converted to plain text, the appropriate parts of the text that require inspection in the subsequent crowd-sourcing step require highlighting (i.e., annotation). We developed a tool utilizing the existing BRAT rapid annotation tool (BRAT)[30] with extensive modifications, for this purpose. BRAT is an annotation tool developed for quickly annotating text for natural language processing (NLP) tasks. The user interface is entirely web-browser based, and can be seen in the supplementary video. Typically, this tool is used to identify relevant parts of the sentence, such as verbs, adverbs, adjectives, etc. It consists of a basic web-based interface, where text is shown and certain words are highlighted with the category labels (adjective, adverb, etc.). As our interests lie primarily in text extraction and not in identifying parts of speech and their relationships, we modified the tool to better suit our needs. We made two substantial modifications to this tool:

(1) We developed a scheme involving the use of regular expressions and keywords to automatically annotate the words we are searching for. For instance, consider parameters such as "laser fluence", "partial pressure of $O_2$", "repetition rate", "substrate termination", "temperature", and "orientation". These can all be identified by their particular units or because they conform to a given form (i.e, (*hkl*) for orientation). Substrate can be determined by comparison with a list of the most common oxide substrates. The text was also mined for the physical characteristics, including thickness, resistivity, transition temperature, magnetization, and polarization, which again are all determined via regular expressions based off the units these quantities are typically measured and reported in. This was accomplished for every paper before it was presented to users. Thus when the user opens a particular paper using the tool, they will see the annotated text of the paper, and a link to the paper in PDF format (see supplementary video).

(2) Apart from annotation of these words, we provide a table that displays all unique materials currently identified, their growth conditions, and their functional properties. Initially the table contains a single entry (titled 'Others'), and each column (which represent the fields in the database) display the various elements found within the paper, for example, that the compound $PbTiO_3$ was found, and the substrate $SrTiO_3$ was found. The task for the user is to fill this table. This is achieved via clicking on the annotated terms (or highlighting text, in case the term was not automatically annotated), confirming the appropriate tag (i.e. field) is selected, and then assigning it to a material grown. Take for example, if we are dealing with $PbTiO_3$ films grown on two substrates, $SrTiO_3$ and $LaAlO_3$, with the substrate temperature 700 °C, with $pO_2$ of 10 mTorr. The user would first select the $PbTiO_3$ (this can be done by clicking on this compound name in the text). This will bring up the annotation window. Then, the user would select 'Material 0' and 'Material 1' in the attributes, and 'compound' in the entity type (though that should be automatically determined, but can be overridden in case it is inaccurate). This will cause two new rows to be added to the table, corresponding to Material 0 and Material 1. The user can then navigate to the points of the text where the temperature of 700 °C is highlighted, click on this text, and again select Material 0 and Material 1 for the attributes. In this way, the table becomes populated.

Our modification of the tool is available at https://github.com/ORNL-DataMatls . To increase efficiency, beside every entry in the table is a line number, and when clicked, will take the user to the point in the text where that entity resides. Additionally, we also offer the user the ability to collapse large sections of text that do not contain any annotations to quickly traverse the document. For completeness, we have provided a video of the tool in use (supplementary). The outcome of this step would be entries in a database. In this case, the database was hosted internally, and the results were stored in an excel spreadsheet, although it can easily be hosted externally in an appropriate database format such as MySQL or MongoDB depending on the needs of the use case. Each user is assigned a subset of all papers, and each paper is seen by at least two users, for verification purposes.

As a paradigmatic example, we considered the case of a class of materials – thin-film epitaxial complex oxides, synthesized by pulsed laser deposition (PLD) – that have been investigated for more than two decades. In this method, a pulsed excimer laser strikes a target of the desired compound to be synthesized, generating a plume of species arriving at a heated substrate. Varying the repetition rate of the laser and duration of deposition allows the quantity of material deposited to be controlled, and careful control of oxygen partial pressure, laser fluence and other growth parameters enables controlled growth of thin-films with unit-cell level precision free of extended defects.[31,32] Given its relative ease, and the fact that PLD can grow extremely high-quality films for a huge variety of oxides, PLD has been a major enabling instrumentation in thin film science in the last three decades.[33] The downside of PLD has been the inability to readily transfer material growth conditions across chambers, requiring local optimization which is time and labor-expensive[34,35]. PLD growth parameters include the choice of substrate (which can result in a strained film, often dramatically influencing properties), the termination of the surface of the substrate, the crystallographic orientation and any deviation from high symmetry orientations of the substrate, the quality of the surface of the substrate (topography roughness, termination), the growth temperature, the partial pressure of oxygen during the growth, and laser parameters such as fluence and repetition rates, and the distance between the target and the substrate. These represent the nominal variables that can be altered by researchers. In practice, other factors may also play a role, such as e.g. the uniformity of heating of the substrate, target density, etc. that are difficult to control for or simply unknown.

**Data Visualization**

An example of some database entries is shown in Table 1, but the whole set can be found elsewhere.[25] The full table contains ~500 entries. We plot the growth conditions (growth temperature and pressure) for four studied compounds in **Figure 2**, with 2D Gaussian fits to the data shown as contours on these plots. The fits allow us to determine the mean and covariance if the data follows a multivariate Gaussian distribution. Note that the fit would be maximal where the point density is large, i.e. the mean growth conditions. This allows easy determination of the range of the growth window for the different compounds. For instance, it is clear that the center for the $BiFeO_3$ is at lower pressures than for $PbZr_xTi_{1-x}O_3$ or the manganites, and it is also somewhat better defined.

Next, we turn to more detailed visualizations for each compound. Shown in **Figure 3** are entries pertaining to the growth of $BiFeO_3$ (BFO) thin films, plotted in scatter plots in Fig. 3(a-d). In Fig. 3(a), the $O_2$ pressure and growth temperatures are plotted against the film thickness, with different colors for different substrates. Apparently, the most commonly used substrate is $SrTiO_3$, though little correlation appears with the growth conditions and the substrate type. More importantly, the scatter plot in Fig. 3(a) shows that the $O_2$ pressure is quite low, mostly below 0.1mbar. There is a slightly larger spread in the reported growth temperatures, mostly between 650 °C and 850 °C. Given that the uncertainties for each point are unknown, there are two possible explanations for the observed scatter: one is that these films can indeed be grown under such varied conditions; alternatively, it could be that the growth window is relatively narrow, but the reporting varies due to e.g. ineffective calibration of pyrometer/heater.

Nonetheless, the results can be compared with respect to first principles calculations of the thermodynamic stability window for the growth of $BiFeO_3$, calculated by Heifets et al.[36] and reproduced in **Figure 5**(a). The graph in Fig. 5(a,b) show the temperatures and oxygen partial pressures under which $BiFeO_3$ is thermodynamically stable, and is reflected by the shaded green region. Other related compounds are preferred at e.g. higher partial pressure of $O_2$ (typically $Bi_2O_3$), and is well-known in the literature. Thus, the narrow stability window predicted by first-principles is well-seen in the results in Fig. 3(a) and Fig. 2(a). It further allows more concrete input for improvement to first-principles databases, given that the growth conditions can now be considered with the associated variance, although the issue of errant entries or unknown calibrations needs addressing. In Fig. 3(b), we plot the film thickness with polarization and growth temperature, but no clear trends are obvious. Due to scaling of ferroelectric thin films, one generally expects to see a decrease in polarization with thickness[37]; however, it may be that because it is in general quite difficult to measure the polarization for ultra-thin films (due to current leakage), these are simply not reported, precluding the ability to determine trends. Also of interest to PLD growers is the laser fluence, which is seen in Fig. 3(c). Apparently, the majority of growers use fluences of under 2J/cm$^2$, and no clear relationship with the growth temperature is found. More interestingly, when the polarization is plotted against the growth pressure (Fig. 3(d)), it appears that there is higher chance of achieving a large polarization when the $O_2$ pressure is somewhat higher. Note the large number of points where no polarization is reported. At first glance, this may be suggested to be due to oxygen vacancies, which would be preferred under lower pressure, although first-principles calculations[38] suggest that the polarization in BFO is robust against oxygen vacancies, so the actual reasons are likely more complex.

We further explored the results for another common ferroelectric, Pb $(Zr_xTi_{1-x})O_3$ (PZT), with results shown in **Figure 4**. Although there are comparatively less statistics available, the $O_2$ pressure window is slightly higher for PZT than for BFO.[39] However, first-principles studies suggest that the surface stability is heavily determined by pO$_2$, as shown by Garrity et al.[40] and reproduced in Fig. 5(c,d). Their calculations show that in the experimentally realizable range of chemical potentials, the surface PbO layer contains a half monolayer of oxygen vacancies for all but a very narrow window of O chemical potentials. If this is correct, then one may conclude that growth of stoichiometric surface layers is not of importance for the majority of the literature (and indeed, should not be of much importance when optimizing for spontaneous polarization, etc.). As with BFO, we may also observe the laser fluence and its relationship with the growth temperature (Fig. 4(c)). With the limited statistics available, it appears that higher fluence is only weakly correlated with lower growth temperatures. This likely derives from the fact that the laser fluence may be considered to be the main factor behind the stoichiometric transfer of material from target to substrate via the plume, i.e. it is somewhat orthogonal to the temperature. The dependence of stoichiometry on laser fluence is well known; for instance, Ohnishi et al.[41] have shown that even small deviations from a critical fluence substantially affects defect structure in the grown film.

Next, we turn to studies of the manganites, specifically La$_{1-x}$(Ca,Sr)$_x$MnO$_3$ (LCMO/ LSMO) thin films. Note that most of the compositions for LSMO/LCMO are on the order of x=0.30-0.40, so we have grouped all the divalent cation compositions into the same parent dataset for illustration and analysis purposes. The results are seen in **Figure 6** with scatter plots. Most reported fluences are below 2 J/cm$^2$, but higher than 1 J/cm$^2$. The reported pressured vary substantially, suggesting that the manganites have a large growth window, especially compared with BFO. There does not appear to be any substantial difference for the reported conditions for either the Ca or Sr case. We have also extracted the transition temperature for each grown manganite (where reported), and is shown in Fig. 6(b,d). Interestingly, there appears to be some degree of correlation in these plots.

**Correlations and Machine Learning Classifier**

Ideally, one may wish to predict the T$_C$ of a given manganite given a set of growth conditions. To determine whether such an analysis is feasible, we may first attempt canonical correlation analysis (CCA)[42] for these two datasets, i.e. a matrix of size *n* x 3, for the growth conditions (laser fluence, temperature and pO$_2$), and another of size *n* x 1 for T$_C$. CCA allows to group variables in matrices such that an optimal correlation is found between the two sets. It returns two sets of canonical coefficients (weights) which reflect differences in the contribution of the different features to the canonical correlation. The CCA *x* and *y* scores, which are obtained by multiplying the standardized original data by the canonical weights matrix, are plotted in **Figure 7**. In an ideal case, this graph would appear linear (implying a correlation coefficient of 1), but the overall correlation coefficient is around 0.6, i.e. rather weak. Nonetheless, it does suggest that there is some dependency, and further, that larger databases may improve on these results.

Given the weak correlation, the use of regression analysis would appear unwarranted; however, one may attempt a more basic classification challenge: given a set of growth conditions, is it possible to determine the likelihood of achieving a low, medium or high $T_C$? We proceeded to classify each $T_c$ value as being high (>300K), medium (200-300K) or low (<200K), and attempted to train a basic machine learning classifier. We utilized a 50/50 train/test split for this task, and trained a decision tree classifier[43], which is a supervised learning algorithm that classifies based on learning decision rules inferred from the input features. The resultant decision tree can be easily visualized as a flowchart, as in **Figure 8**(a). Similarly, it is possible to plot the decision tree surface for any two variables, as in Fig. 8(b), although this can be misleading as in fact the decision tree surface is generally multidimensional (in our case, it is 3-dimensional), so care must be taken. The raw data is plotted as filled circles, while the colors represent the different classes. Apparently, lower temperatures correspond to higher chance of classification in the 'high $T_c$' range. We found that the accuracy of this classification was limited, to no better than ~70%. Nonetheless, it shows the promise of using literature data towards understanding complex, nonlinear interactions that are otherwise difficult to determine by any one experimental group, and validates the purpose of this study.

**Discussion**

The tools provided in this manuscript represent a first-attempt towards a tackling a significant challenge in the synthesis of complex oxides. However, this approach is universal, and can be applied equally for other synthesis approaches. Although in our case the correlations were somewhat low, these may simply be a result of lack of enough data, and opportunities exist for rapid expansion of the given databases. The first task in this endeavor would be to allow research groups from around the world to add their own entries, based on the depositions performed in their own individual labs. As a related matter, laboratory notebooks and logbooks are a treasure-trove, because logbooks not only report successes, but importantly, failures – a recent notable example published in *Nature* shows through the use of all available information on synthesis, including failures, it was possible to predict reaction outcomes for templated vanadium selenites.[44] Indeed, failures are critical to understand[45] (see also, the Dark Reactions Project[15]), for they will not generally be observed in the agglomerated literature (negative results are almost never published), and thus, it is difficult to ascertain why certain regions of the parameter space are unexplored. For optimal experimental design and forward modeling, such information is critical, and indeed just as important as positive results for understanding synthesis pathways. Our approach allows to leverage this information, e.g. with the addition of a boolean field for 'unsuccessful synthesis', and a note on why (e.g., secondary phase, 3D growth, etc.).

At present, much of the tagging and building of the database entries still requires human efforts. Though there have been substantial improvements in natural language processing capabilities via approaches such as deep learning[46] (and could potentially be applied here as well), much of the important information in each paper is buried within individual figures. In future, determination of the relevant figures, possibly using suitable classification algorithms[47], followed by digitization and extraction of the important values (such as remnant polarization or $T_c$) would be a critical step towards automatic database generation.

As a related point, the issue of chemical heterogeneities of samples and variability from different processing conditions, and the impact on functional properties and final device performance, is an important one to address.[48] Necessarily, it implies the need to aggregate the information from different groups, and then data mining methods[49] to determine the relevant processing parameters. Databases of experimental results such as the one constructed here are well suited towards this application, as they can help to answer the question of sample variability quantitatively.

Of course, there are other avenues for the use of this type of database. For instance, if variability is found in growth conditions, and there exists particular clusters, one may ask as to the source of this behavior. It may be that most groups cluster around a particular set of growth conditions, but one or two individual groups are outliers. If the functional properties of these groups are markedly different, then there is basis for further investigation. As an example, see table from a recent review of T-$BiFeO_3$.[50] This table was compiled manually, and took several weeks.[51] Note though that although possible in the case for a single compound that has only been studied for ~5 years, it would be practically impossible to do for ~5000 papers or materials that have been studied extensively for 20+ years, and speaks to the need for the databases like the one reported here.

Perhaps more importantly, the database can be used as a starting point to synthesize new, related compounds. For instance, if synthesizing a particular perovskite manganite with one type of divalent cation, then a lookup of the table will immediately guide the approximate bounds of the parameter space to explore for the new compound (after correlating to thermodynamics), saving significant time. Additionally, mining of the data can provide latent connections, for example, between physical properties and electrochemistry for the manganites. In addition, the fact that we can know the temperature and partial pressure of oxygen in the growth provides bounds on the space of the T-p-c diagram in which the material resides, and therefore allows estimation of the number of oxygen vacancies, the types of surface reconstructions, or even if the system is thermodynamically stable.

Similar to previous works with first-principles databases (examples including Materials Project[52], AFLOWLIB[53], JARVIS[54], NRELMatDB[55]), this experimental database will be ripe for future investigations by more complex learning algorithms such has been done by e.g. Curtarolo et al.[11,56,57], which could provide fresh insights into the physics of these systems. Moreover the properties can be matched with predictions from thermodynamic and first principles calculations which can provide information on the factors controlling the structure-property relationships, which are difficult (if not impossible) to discern from individual experiments where variabilities are simply too large. Finally, we expect these methods to be applicable to a wide range of materials classes and growth types. For example, it could be used for battery electrodes to determine which dopants and synthesis routes are linked with the highest ionic conductivity, lower degradation, etc.

In summary, we present a method to semi-automatically mine the extant literature for information pertaining to materials synthesis. We apply this to an example of pulsed laser deposition of oxide thin films, and use crowd-sourcing in conjunction with open-source developed software to tag processing condition and physical property information on hundreds of deposited thin films in the past three decades. We find the growth windows for a number of compounds, discuss these in light of their thermodynamic stability windows, and attempt to correlate their physical property metrics with their processing conditions. These experimental databases offer the potential to dramatically alter the material synthesis landscape, allowing tighter feedback with computational modeling, exploring latent connections, and incorporating data from failed or unpublished studies, greatly expanding the chance of success of initiative such as the Materials Genome Initiative. The tools provided here show the pathway toward a data and community-driven approach towards understanding material synthesis and associated pathways, and will comprise a significant tool in the goal of an artificial intelligence for materials discovery.

**Supplementary Material**

Please see supplementary material for the full Jupyter notebook, along with database file and video of the annotation tool in use. The notebook contains all the code necessary to conduct the analysis and generate the plots used in this manuscript.

**Acknowledgements**

We acknowledge fruitful discussions with H.-N. Lee (ORNL) and J.-C Yang (National Cheng Kung University). This research was sponsored by the Laboratory Directed Research and Development Program of Oak Ridge National Laboratory, managed by UT-Battelle, LLC, for the US Department of Energy (DOE). A portion of this research as sponsored by the U.S. Department of Energy, Office of Science, Basic Energy Sciences, Materials Sciences and Engineering Division (SVK). A portion of this research was conducted at the Center for Nanophase Materials Sciences, which is a US DOE Office of Science User Facility.

**Table I: Some entries from the compiled database**

| Compound | Substrate | Thickness (nm) | Growth Temperature (°C) | Repetition Rate (HZ) | $pO_2$ (mbar) | $T_c$ (K) | Fluence (J/cm²) | Remnant Polarization (μC/cm²) | doi | Reference |
|---|---|---|---|---|---|---|---|---|---|---|
| $La_{2/3}Sr_{1/3}MnO_3$ | (001) $SrTiO_3$ | | 660 | 3.3 | 0.2260 | 305 | | | 10.1063/1.1949727 | Ref[58] |
| $Pb(Zr_{0.2}Ti_{0.8})O_3$ | SrTiO3 | 50 | 600 | 5 | 0.2000 | | 0.5 | 80 | 10.1016/j.apsusc.2005.07.149 | Ref[59] |
| $BiFeO_3$ | (111) $LaAlO_3$ | 200 | 650 | | 0.0267 | | 2 | | 10.1103/PhysRevB.73.092408 | Ref[60] |
| $BiFeO_3$ | (110) $SrTiO_3$ | 300 | 610 | 2 | 0.1000 | | 1.5 | | 10.1063/1.4902113 | Ref[61] |

*Figure Captions*

**Figure 1: Workflow for text mining of relevant papers from the literature.** A list of documents is collected from a keyword search of relevant terms. The documents are then downloaded and converted to plain text. They are then passed through the annotation tool, which automatically parses the text and highlights key terms pertaining to processing conditions and physical properties. Crowd-sourcing is then used to sift through each paper to match individual films grown with their physical properties, automatically updating a centralized database. Subsequent scripts allow normalizing the database, for further analysis.

**Figure 2: Growth conditions for four compounds studied.** In **(a-d)** we plot the growth conditions (substrate temperature and $O_2$ pressure) for (a) $BiFeO_3$, (b) $PbZr_xTi_{1-x}O_3$, (c) $La_{1-x}Ca_xMnO_3$ and (d) $La_{1-x}Sr_xMnO_3$ thin films. Each point represents a single film deposition.

**Figure 3: Visualization for results for $BiFeO_3$.** **(a)** Growth conditions plotted against film thickness, with the colors indicating different substrates (see legend inset). **(b)** Film thickness and growth temperature plotted against the remnant polarization. **(c)** Relationship between fluence and growth temperatures, and **(d)** dependence of the polarization on the $O_2$ pressure during growth. Note the many points in (d) where no polarization was measured (default value is 0).

**Figure 4: Visualization of results for $PbZr_xTi_{1-x}O_3$.** **(a)** Growth conditions plotted against film thickness, with the colors indicating different substrates (see legend inset). **(b)** Film thickness and growth temperature plotted against the remnant polarization. **(c)** Relationship between fluence and growth temperatures, and **(d)** Dependence of the polarization on the $O_2$ pressure during growth. Note the many points in (c) where no fluence was given (default value is 0).

**Figure 5: First-principles calculations for the stability window for $BiFeO_3$ and the surface of $PbTiO_3$ (a-b)** Thermodynamic stability window for $BiFeO_3$, shown as the green region in (a). Different lines correspond to distinct compounds. The chart in (b) can be used to convert the chemical potential of oxygen to a partial pressure, for different temperatures. In **(c-d)**, the surface stability for $PbTiO_3$ surfaces, for (c) paraelectric (d) negatively poled surfaces is reproduced. (a-b) is Reprinted (adapted) with permission from Heifets et al.[36] Copyright (2015) American Chemical Society. (c-d) is Reprinted with permission from Garrity et al.[40], Copyright (2013) by the American Physical Society.

**Figure 6: Visualization of results for $La_{1-x}Sr_xMnO_3$ and $La_{1-x}Ca_xMnO_3$.** Scatter plot of laser fluence and growth temperature for **(a)** LSMO and **(c)** LCMO. Dependence of the magnetic transition temperature, $T_c$, on the $O_2$ pressure during growth for **(b)** LSMO and **(d)** LCMO.

**Figure 7: Canonical Correlation Analysis for $T_c$ of manganites as a function of three different growth variables.** The CCA *x* and *y* scores are obtained by multiplying the standardized original

data by the canonical weights matrix. In the ideally correlated case, this graph would exhibit an exact linear relationship with slope 1 (red dashed line).

**Figure 8: Decision Tree Classifier for $T_c$, given some growth conditions.** (a) Decision tree visualization as a chart. (b) Visualizing the decision tree surface in two dimensions. Data is shown as filled circles, and colored based on the classification into the different critical temperatures. The colored regions correspond to the low, medium and high $T_c$ classification regions of the parameter space (light blue = low $T_c$, darker blue = medium $T_c$, red = high $T_c$).

*Supplementary Material*

**Data Mining for better material synthesis:**

**the case of pulsed laser deposition of complex oxides**

Steven R. Young, Artem Maksov, Maxim Ziatdinov, Ye Cao, Matthew Burch,

Janakiraman Balachandran, Linglong Li, Suhas Somnath, Robert M. Patton,

Sergei V. Kalinin and Rama K. Vasudevan

**Supplementary Video 1 – Video of BRAT Annotation tool in use**

This video shows a demonstration of the modified BRAT rapid annotation tool. A paper is loaded, and entries are added to the table within the tool.

**Supplementary Jupyter Notebook and Data Files**

The full Jupyter notebook that was used to generate the figures and conduct the analysis is provided along with the database as a comma-separated values file.

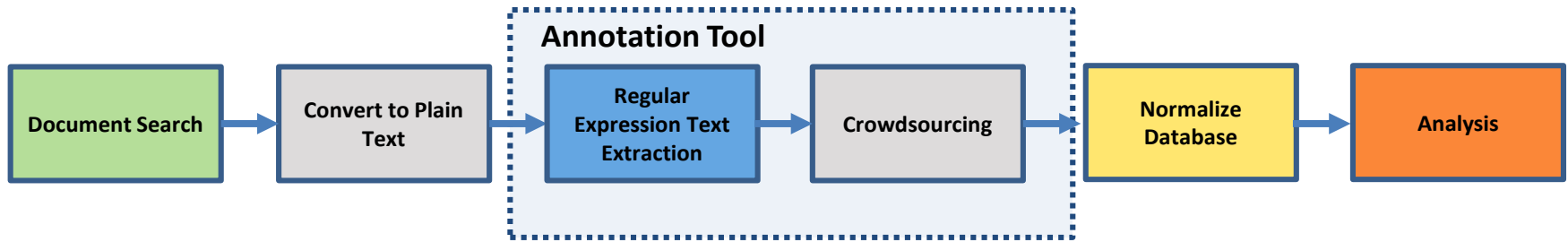

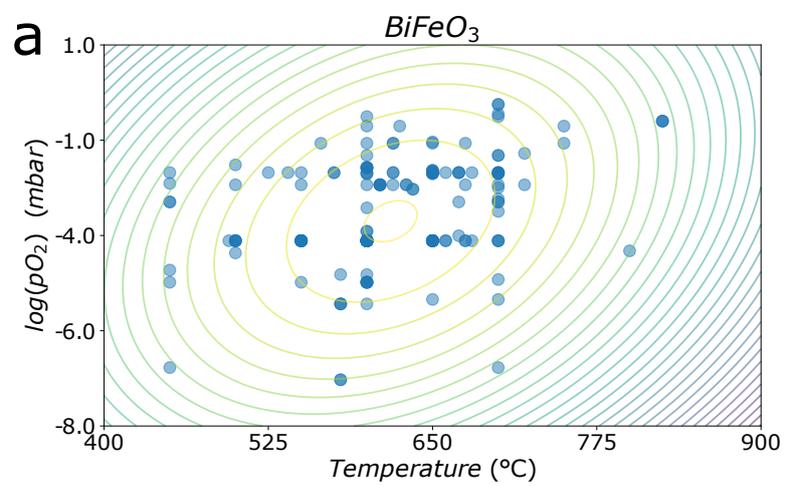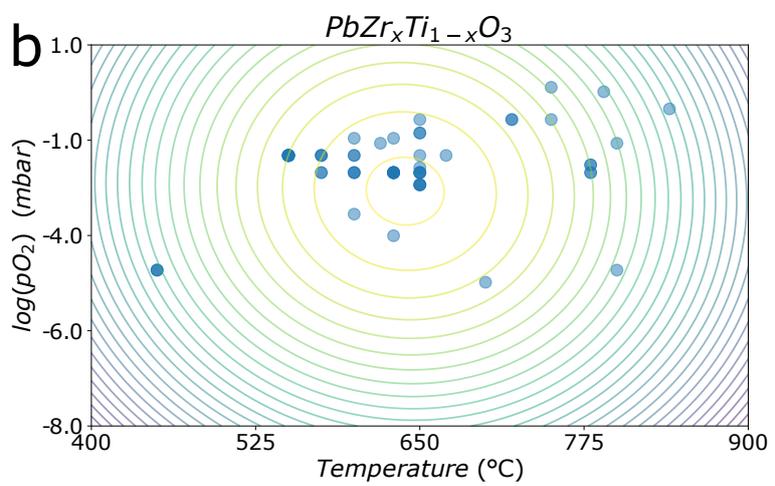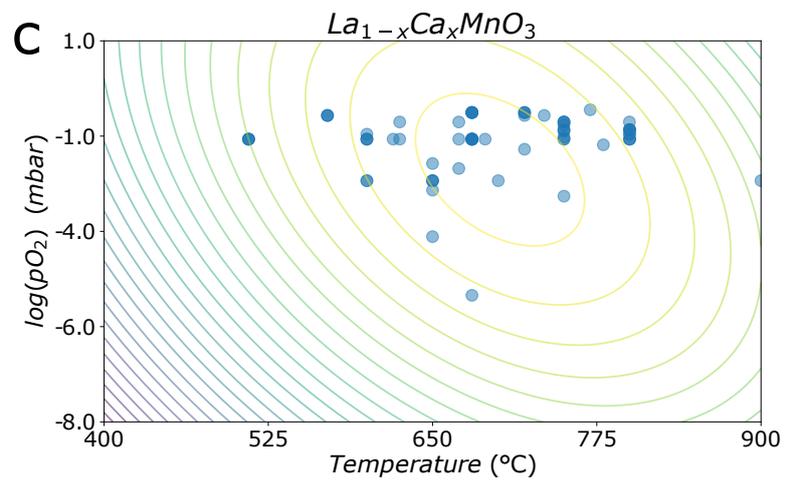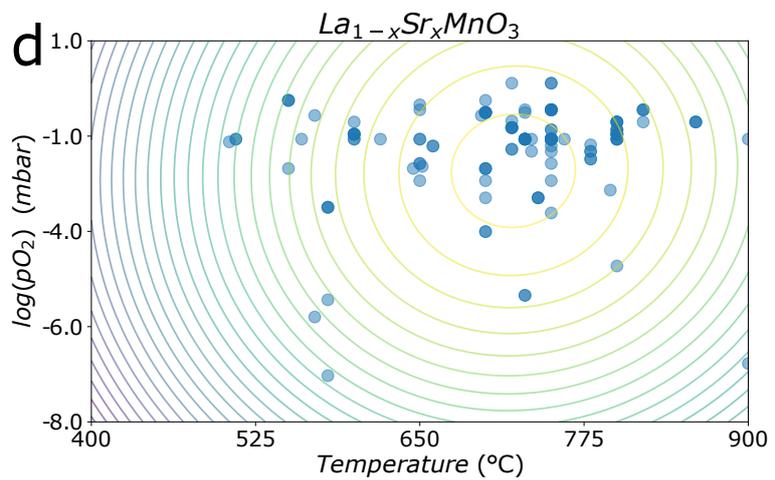

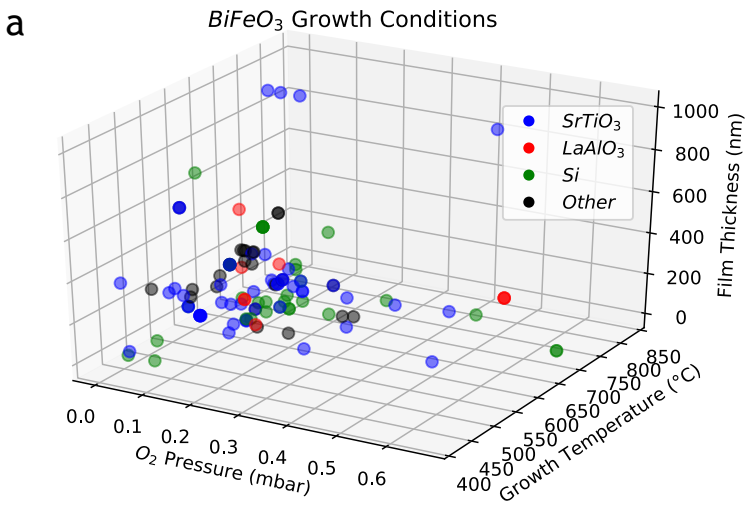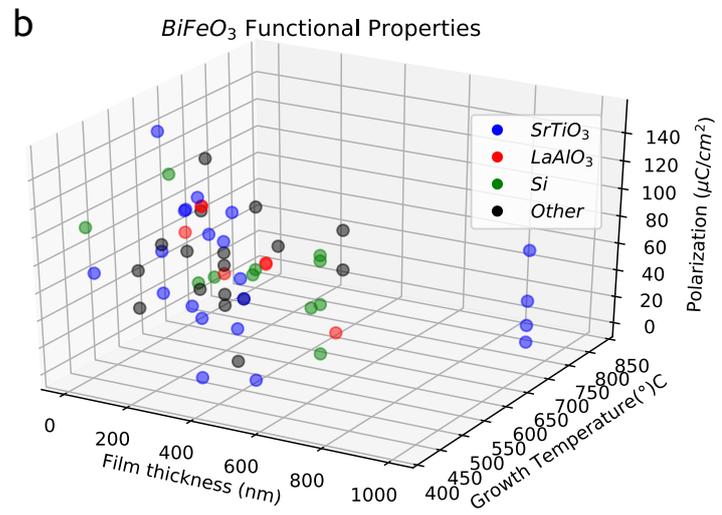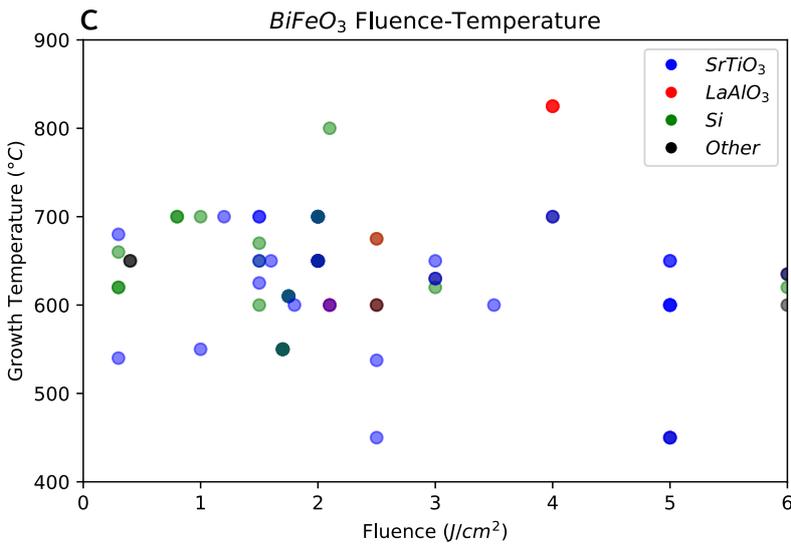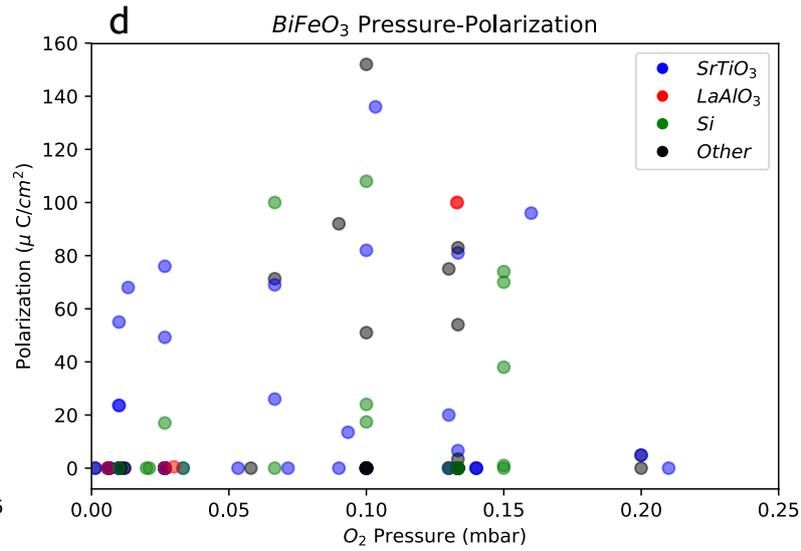

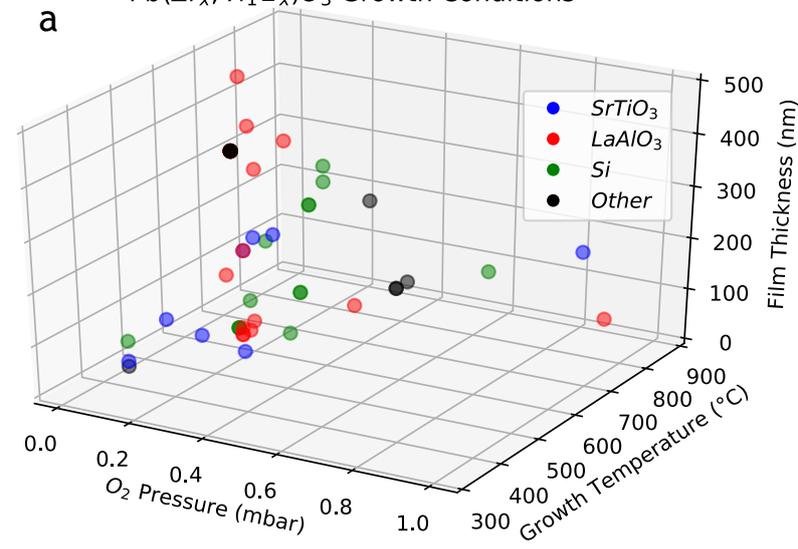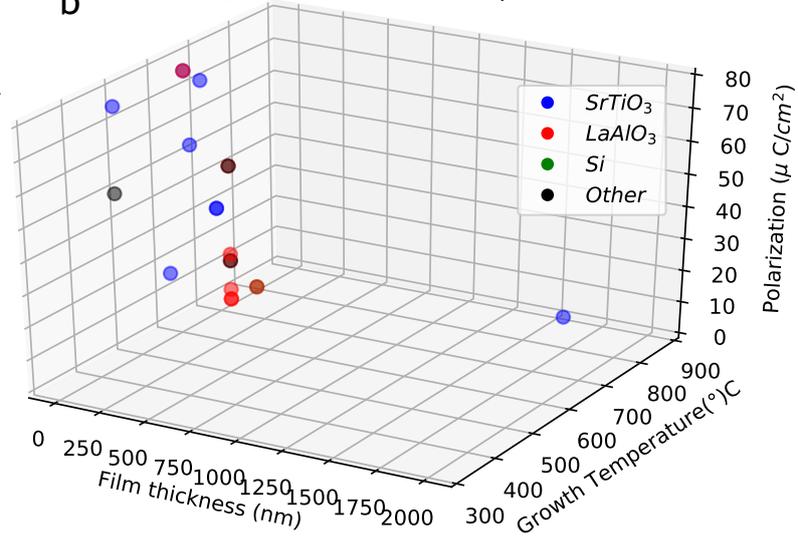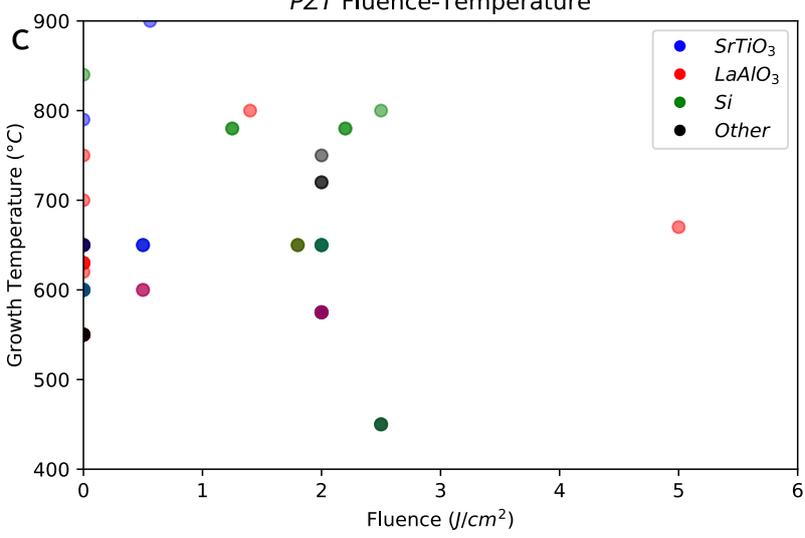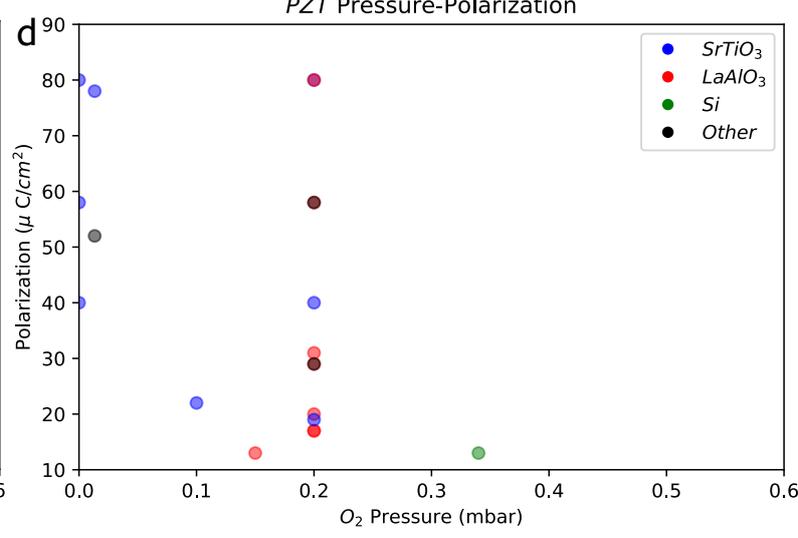

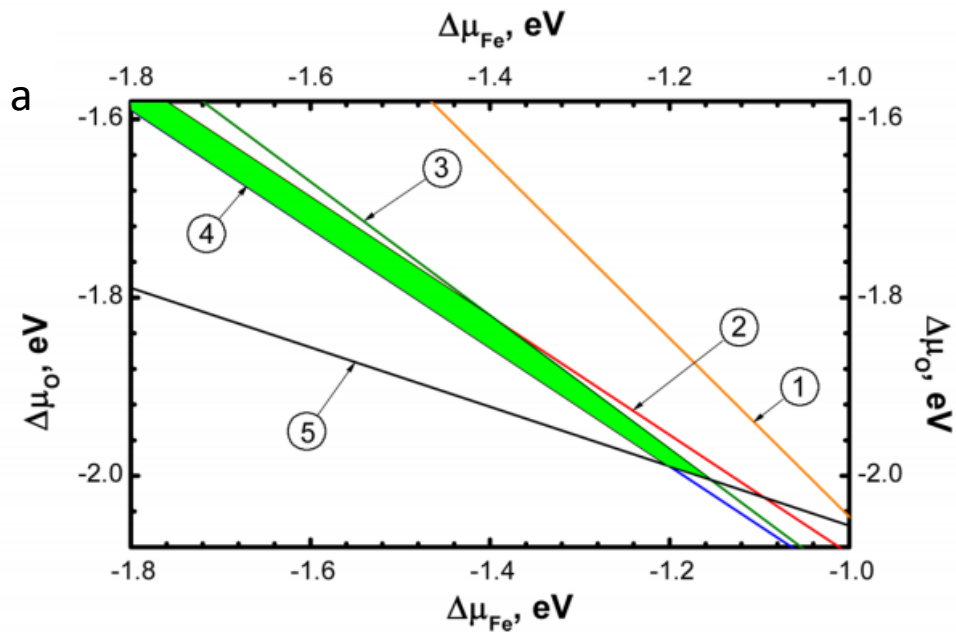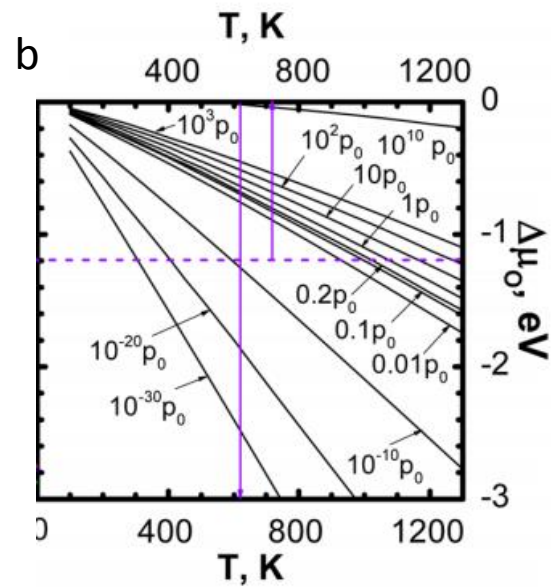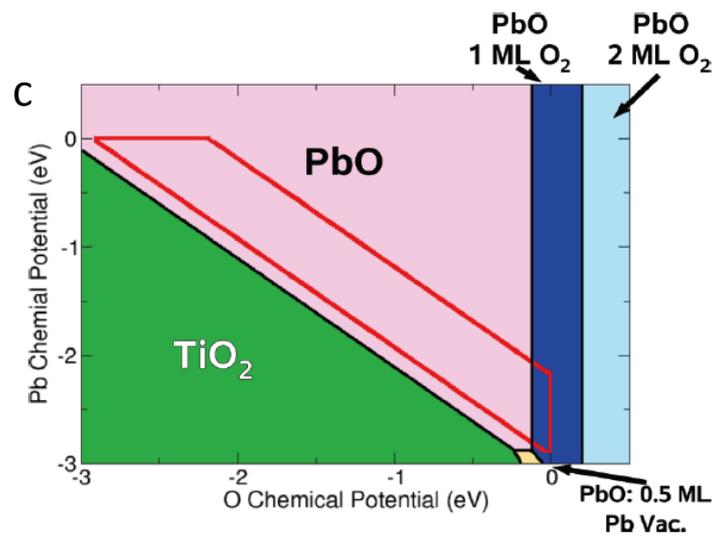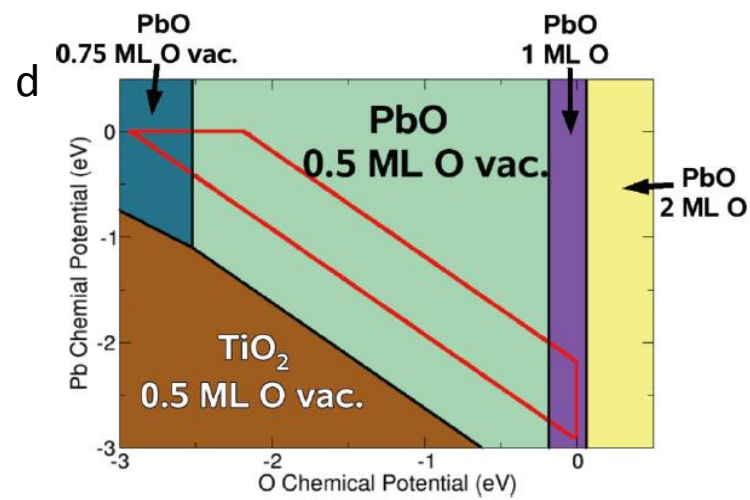

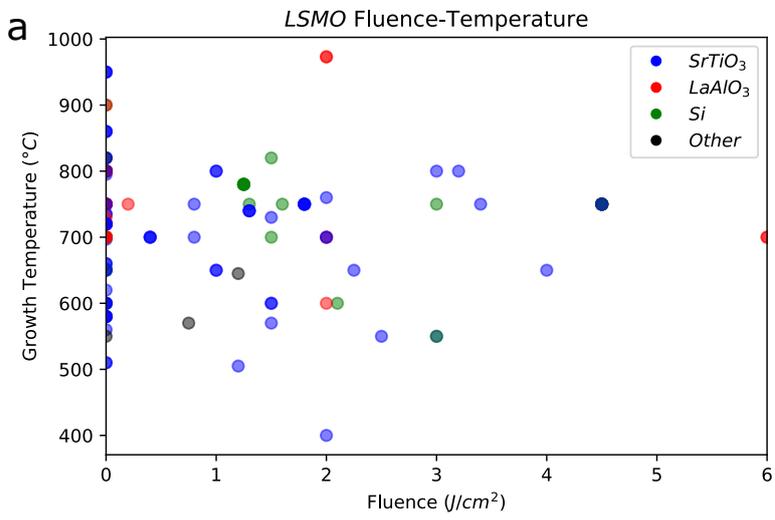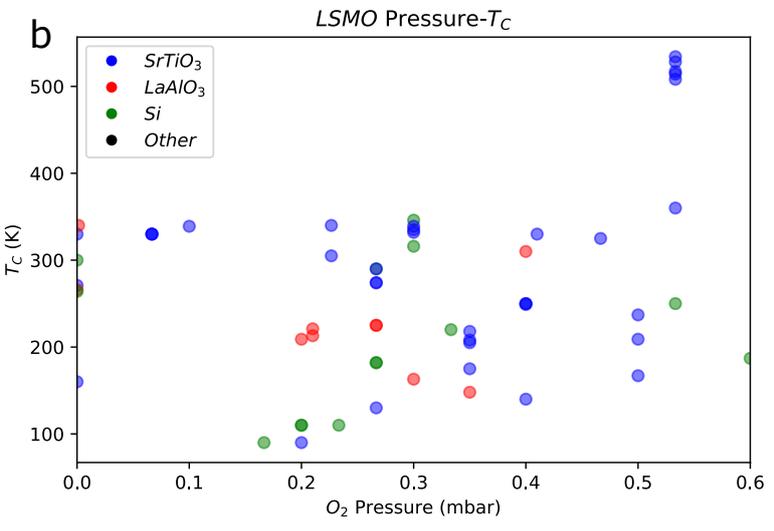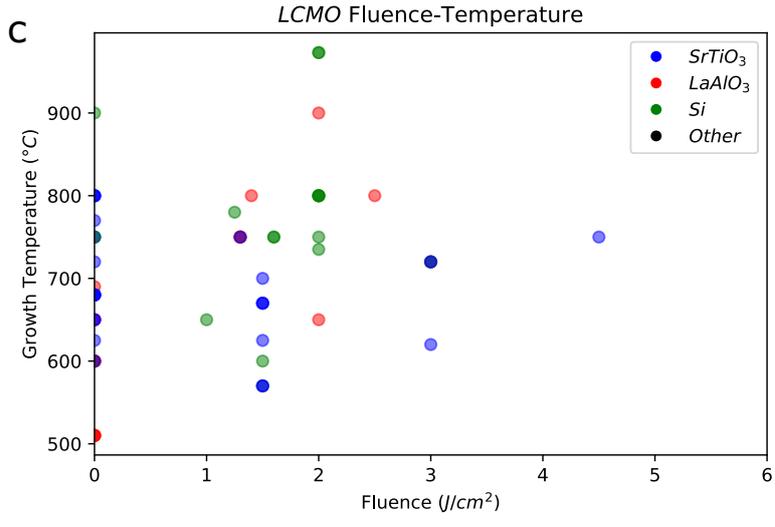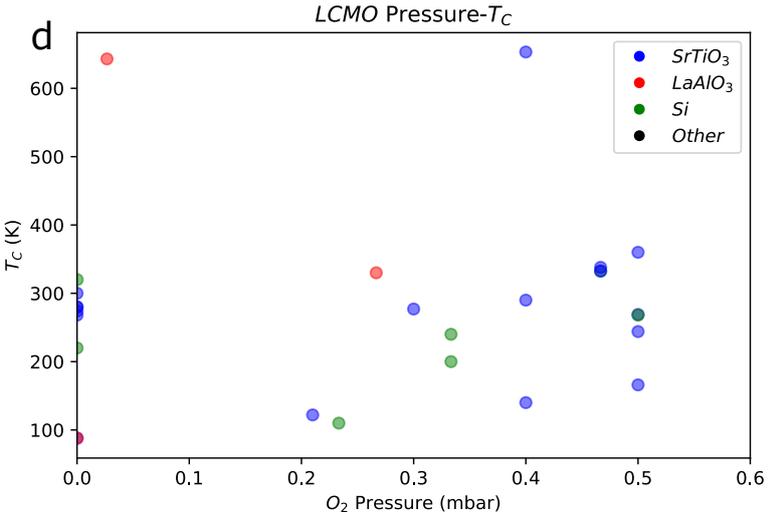

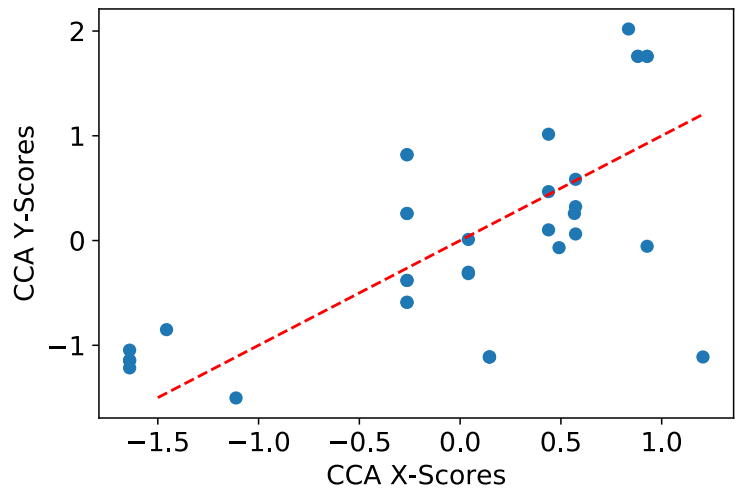

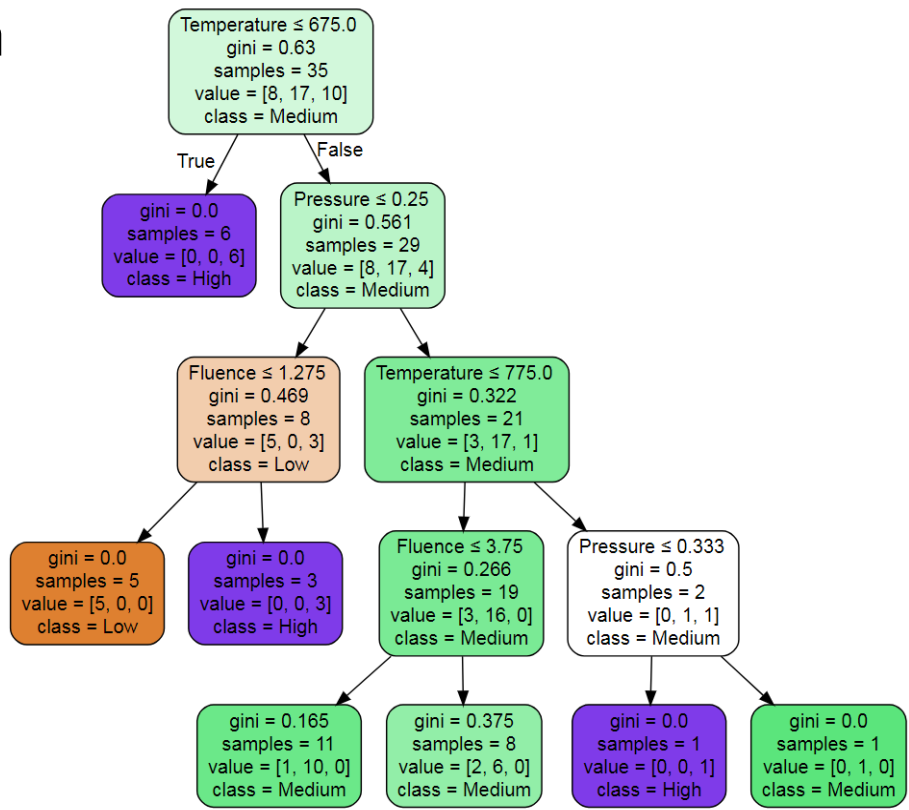
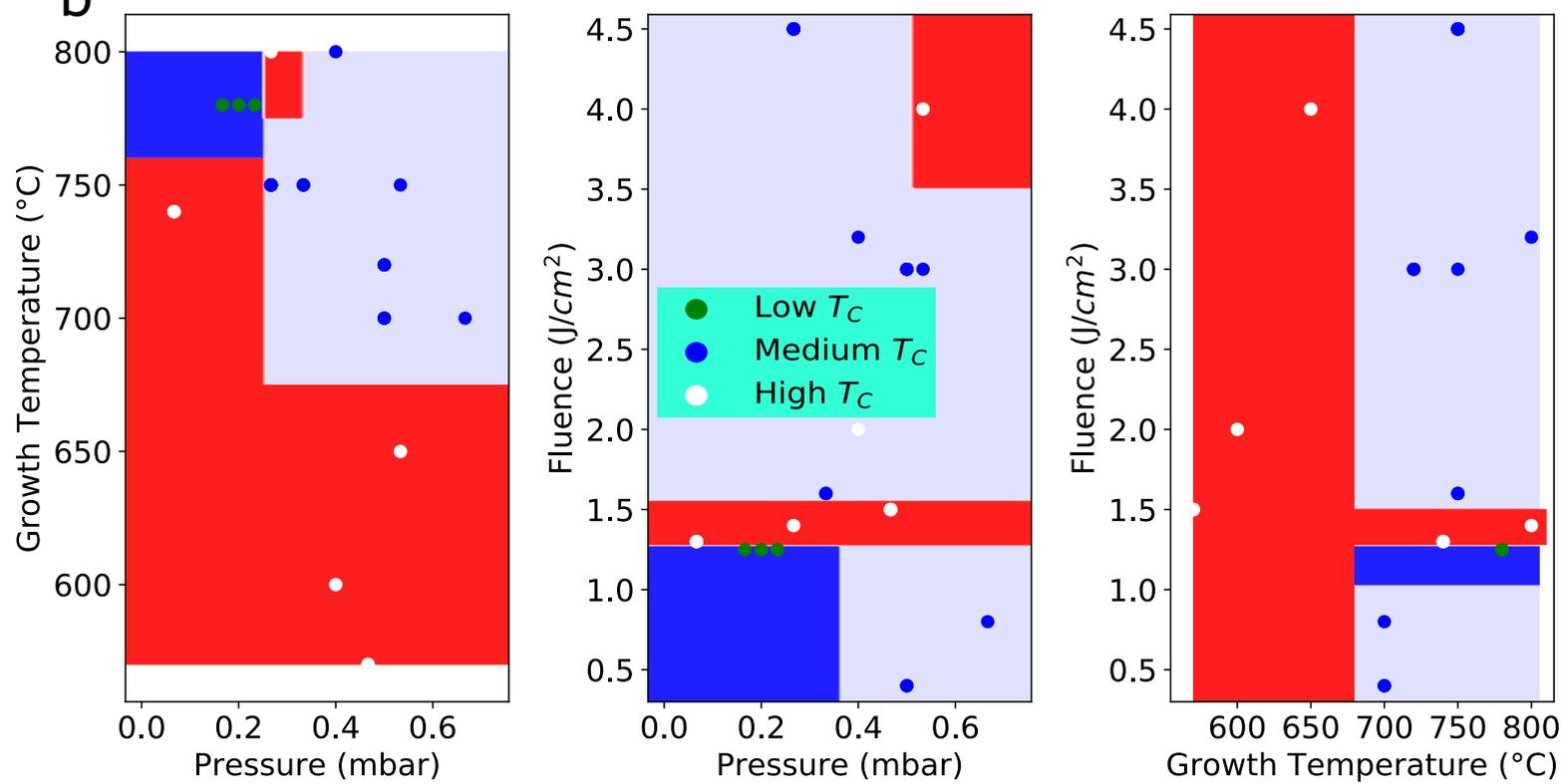